\documentclass[10pt]{article}
\usepackage{amsmath}
\usepackage{amssymb}
\usepackage{epsfig}
\begin{document}
\numberwithin{equation}{section}
\numberwithin{table}{section}
\numberwithin{figure}{section}

\begin{table}[t]
  \begin{flushright}
     OUTP-04-03P    \\
     hep-th/0401114 \\
     January 2004
  \end{flushright}
\end{table}

\title{$ SO(10) $ heterotic M-theory vacua \footnote{Invited talk
presented at the String Phenomenology 2003 Workshop, IPPP, Durham UK, 29 
July - 4 August 2003.}}

\author{Richard S. Garavuso  
\\
\emph{\footnotesize Theoretical Physics Department, 
University of Oxford, Oxford \ OX1 3NP, UK}
\\
\footnotesize{garavuso@thphys.ox.ac.uk}
}

\date{ }

\maketitle

\begin{abstract}

This talk adapts the available formalism to study a class of
heterotic M-theory vacua with $ SO(10) $ grand unification group.
Compactification to four dimensions with  $ \mathcal{N} = 1 $
supersymmetry is achieved on a torus fibered Calabi-Yau 3-fold
$ \mathbf{Z} = \mathbf{X} / \tau_{\mathbf{X}} $ with first homotopy group 
$ \pi_{1}(\mathbf{Z}) = \mathbb{Z}_{2} $. 
Here $ \mathbf{X} $ is an elliptically fibered Calabi-Yau 3-fold which
admits two global sections and $ \tau_{\mathbf{X}} $ is a freely acting
involution on $ \mathbf{X} $. The vacua in this class have net number of
three generations of chiral fermions in the observable sector and
may contain M5-branes in the bulk space which wrap holomorphic
curves in $ \mathbf{Z} $.  Vacua with nonvanishing and vanishing instanton
charges in the observable sector are considered.  The latter case
corresponds to potentially viable matter Yukawa couplings.  Since
$ \pi_{1}(\mathbf{Z}) = \mathbb{Z}_{2} $, the grand unification group can
be broken with $ \mathbb{Z}_{2} $ Wilson lines.

The motivation is to use the above formalism to extend realistic
free-fermionic models to the nonperturbative regime.
The correspondence between these models and
$ \mathbb{Z}_{2} \times \mathbb{Z}_{2} $ orbifold compactification of the 
weakly coupled 10-dimensional heterotic string identifies associated
Calabi-Yau 3-folds which possess the structure of the above $ \mathbf{Z} $
and $ \mathbf{X} $. A nonperturbative extension of
the top quark Yukawa coupling is discussed.

\end{abstract}

%%%%%%%%%%%%%%%%%%%%%%%%%%%%%%%%%%%%%%%%%%%%%%%%%%%%%%%%%%%%%%%%%%%%%%%%%
%%%%%%%%%%%%%%%%%%%%%%%%%%%%%%%%%%%%%%%%%%%%%%%%%%%%%%%%%%%%%%%%%%%%%%%%%
%%%%%%%%%%%%%%%%%%%%%%%%%%%%%%%%%%%%%%%%%%%%%%%%%%%%%%%%%%%%%%%%%%%%%%%%%
\section{\label{Intro}Introduction} 

M-theory on the orbifold $ \mathbf{S}^{1} / \mathbb{Z}_{2} $
is believed to describe the strong coupling limit of the 
$ E_{8} \times E_{8} $ heterotic string \cite{HorWit:Het}.  
At low energy, this theory is described by 11-dimensional supergravity
coupled to a chiral $ \mathcal{N} = 1 $, $ E_{8} $ Yang-Mills
supermultiplet on each of the two 10-dimensional orbifold fixed planes
\cite{HorWit:Ele}.  This low energy description is known as
\emph{Ho\v{r}ava-Witten theory}.

Compactifications of Ho\v{r}ava-Witten theory leading to 
$ \mathcal{N} = 1 $ supersymmetry in four dimensions \cite{Wit:Str}
are (to lowest order) based on the spacetime structure
\begin{equation}
\mathbf{M}^{11} = \mathbf{M}^{4} \times \mathbf{S}^{1}/\mathbb{Z}_{2}
\times \boldsymbol{\mathcal{X}},
\end{equation}
where $ \mathbf{M}^{4} $ is 4-dimensional Minkowski space and 
$ \boldsymbol{\mathcal{X}} $ is a Calabi-Yau 3-fold.
Generally, on a given fixed plane, some subgroup $ G $ of the $ E_{8} $
symmetry survives on $ \boldsymbol{\mathcal{X}} $. $ E_{8} $ is broken to
$ G \times H
$, where the grand unification group $ H $ is the commutant subgroup of 
$ G $ in $ E_{8} $. The `standard embedding', in which $ G = SU(3) $ is
identified with the spin connection of $ \boldsymbol{\mathcal{X}} $,
corresponds to 
$ H = E_{6} $.  Vacua with nonstandard embeddings may contain M5-branes 
\cite{Wit:Str,LukOvrWal:Non} in the bulk space. One refers
to Ho\v{r}ava-Witten theory compactified to lower dimensions with
arbitrary gauge vacua as \emph{heterotic M-theory}.

Donagi, Ovrut, Pantev and Waldram \cite{DonOvrPanWal}
presented a class of heterotic M-theory vacua with $ H = SU(5) $ grand
unification group.  The vacua in this class have net number of generations 
$ N_{\textrm{gen}} = 3 $ of chiral fermions in the observable sector and
may contain M5-branes in the bulk space at specific points in the
orbifold direction.  Compactification to four dimensions with 
$ \mathcal{N} = 1 $ supersymmetry is achieved on a \emph{torus} fibered
Calabi-Yau 3-fold
\begin{equation}
\label{Z}
\mathbf{Z}  = \mathbf{X} / \tau_{\mathbf{X}}
\end{equation}
with first homotopy group
\begin{equation}
\pi_{1}(\mathbf{Z}) = \mathbb{Z}_{2}.
\end{equation}
Here $ \mathbf{X} \stackrel{\pi}{\rightarrow} \mathbf{B} $ is a torus
fibered Calabi-Yau 3-fold which admits two global sections, a zero
section $ \sigma $ and a second section $ \xi $, and $ \tau_{\mathbf{X}} $
is a freely acting involution on $ \mathbf{X} $.  A torus fibered manifold
which admits a zero section is said to be \emph{elliptically} fibered.
Since $ \pi_{1}(\mathbf{Z}) = \mathbb{Z}_{2} $, the $ H = SU(5) $
grand unification group can be broken to the Standard Model gauge group
with a $ \mathbb{Z}_{2} $ Wilson line. The M5-branes are required to 
span $ \mathbf{M}^{4} $ and wrap holomorphic curves in $ \mathbf{Z} $. 

This talk extends the above work by Donagi, Ovrut, Pantev and Waldram by
considering the $ H = SO(10) $ grand unification group.  The $ SO(10) $
embedding of the Standard Model spectrum is supported by experimental
evidence \cite{neutrino} for neutrino masses.  A class of string models
which preserves this embedding are the \emph{realistic free-fermionic
models} \cite{(FSU5),(PS),(SLM),(LRS)}. These $ N_{\textrm{gen}} = 3 $,
4-dimensional string models are constructed using the free-fermionic
formulation \cite{free} of the weakly coupled heterotic string.  
The correspondence \cite{correspond} between these models and
$ \mathbb{Z}_{2} \times \mathbb{Z}_{2} $ orbifold compactification of the
weakly coupled 10-dimensional heterotic string identifies associated
Calabi-Yau 3-folds which possess the structure of the above $ \mathbf{Z} $
and $ \mathbf{X} $.  This allows realistic free-fermionic models to
be studied in the nonperturbative regime.

The new results presented in this talk have been published by
Faraggi, Garavuso and Isidro \cite{FarGarIsi:Non} and
Faraggi and Garavuso \cite{FarGar:Yuk}.  Section \ref{Realistic} reviews
realistic free-fermionic models and their 
$ \mathbb{Z}_{2} \times \mathbb{Z}_{2} $ correspondence, explains the
connection \cite{FarGarIsi:Non} with the Calabi-Yau 3-folds 
$ \mathbf{Z} $ and $ \mathbf{X} $, and offers a proposal to obtain a
nonperturbative extension \cite{FarGar:Yuk} of the top quark Yukawa
coupling calculation \cite{top}.  We will see that, although the precise
geometrical realization of the full $ N_{\textrm{gen}} = 3 $ models is
unknown, this coupling can be computed as a $ \mathbf{27}^{3} $ $ E_{6} $
or $ \mathbf{16} \cdot \mathbf{16} \cdot \mathbf{10} $ $ SO(10) $ coupling
with $ N_{\textrm{gen}} = 24 $.  This leads us to present rules for
constructing heterotic M-theory vacua allowing $ H = E_{6} $ and 
$ SO(10) $ grand unification groups with arbitrary $ N_{\textrm{gen}} $.
Section
\ref{SumRul} summarizes rules \cite{FarGar:Yuk} allowing grand unification
groups $ H = E_{6}$ , $ SO(10) $, and $ SU(5) $, corresponding to 
$ G = SU(n) $ with $ n = 3 $, 4, and 5, respectively.  As discussed by
Arnowitt and Dutta \cite{ArnDut:Yuk}, requiring vanishing instanton
charges in the observable sector yields potentially viable matter Yukawa 
couplings.  $ N_{\textrm{gen}} = 3 $, $ H = SO(10) $ vacua with
nonvanishing \cite{FarGarIsi:Non} and vanishing \cite{FarGar:Yuk}
instanton charges in the observable sector are discussed in Section 
\ref{Vacua}.

%%%%%%%%%%%%%%%%%%%%%%%%%%%%%%%%%%%%%%%%%%%%%%%%%%%%%%%%%%%%%%%%%%%%%%%%%%
%%%%%%%%%%%%%%%%%%%%%%%%%%%%%%%%%%%%%%%%%%%%%%%%%%%%%%%%%%%%%%%%%%%%%%%%%%
\section{\label{Realistic}Realistic free fermionic models}

%%%%%%%%%%%%%%%%%%%%%%%%%%%%%%%%%%%%%%%%%%%%%%%%%%%%%%%%%%%%%%%%%%%%%%%%%%
\subsection{General structure}

A free-fermionic model is generated by a suitable choice of boundary
condition basis vectors (which encode the spin structure of the worldsheet
fermions) and generalized GSO projection coefficients.  The boundary
condition basis vectors associated with the realistic free-fermionic
models are constructed in two stages.  The first stage constructs the NAHE
set \cite{FarNan:Nat} of five basis vectors denoted by 
$\{\mathbf{1},\mathbf{S},\mathbf{b}_{1},\mathbf{b}_{2},\mathbf{b}_{3}\}$.
After generalized GSO projections over the NAHE set, the residual gauge
group is 
\begin{equation*}
SO(10) \times SO(6)^{3} \times E_{8}.  
\end{equation*}
NAHE set models have
$ \mathcal{N} = 1 $ spacetime supersymmetry
and 48 chiral generations in the $ \mathbf{16} $
representation of $ SO(10) $ (16 from each of the sectors 
$ \mathbf{b}_{1} $, $ \mathbf{b}_{2} $, and $ \mathbf{b}_{3} $).
The sectors $ \mathbf{b}_{1} $, $ \mathbf{b}_{2} $, and $ \mathbf{b}_{3} $
correspond to the three twisted sectors of the associated 
$ \mathbb{Z}_{2} \times \mathbb{Z}_{2} $ orbifold.  The second stage of
the construction adds three (or four) basis vectors, typically denoted by
$\{ \boldsymbol{\alpha},\boldsymbol{\beta},\boldsymbol{\gamma},\ldots \}$, 
which correspond to Wilson lines in the associated 
$ \mathbb{Z}_{2} \times \mathbb{Z}_{2} $ orbifold formulation.  These
basis vectors break the $ SO(10) \times SO(6)^{3} \times E_{8} $ gauge
group and reduce the number of chiral generations from 48 to 3 (one from
each of the sectors
$ \mathbf{b}_{1} $, $ \mathbf{b}_{2} $, and $ \mathbf{b}_{3} $).
The $ SO(10) $ symmetry is broken to one of its subgroups. The
flipped $ SU(5) $ \cite{(FSU5)},
Pati-Salam \cite{(PS)},
Standard-like \cite{(SLM)}, and
left-right symmetric \cite{(LRS)}
$ SO(10) $ breaking patterns are shown in Table \ref{SO(10)breaking}.
\begin{table}[t]
$$ \begin{array}{cl}
\textrm{flipped} \ SU(5) & SO(10) \rightarrow SU(5) \times U(1)
\\
\textrm{Pati-Salam}      & SO(10) \rightarrow SO(6) \times SO(4)
\\
\textrm{Standard-like}   &
SO(10) \rightarrow SU(3) \times SU(2) \times U(1)^{2}
\\
\textrm{left-right symmetric} & 
  SO(10) \rightarrow SU(3) \times SU(2)_{L} \times SU(2)_{R} \times U(1)
\end{array} $$
\caption{$ SO(10) $ breaking patterns in realistic free-fermionic models.}
\label{SO(10)breaking}
\end{table}
In the former two cases, an additional 
$ \mathbf{16} $ and $ \mathbf{\overline{16}} $
representation of $ SO(10) $ is obtained from the set
$\{\boldsymbol{\alpha},\boldsymbol{\beta},\boldsymbol{\gamma},\ldots\}$.
Similarly, the hidden $ E_{8} $ is broken to one of its
subgroups.  The flavor $ SO(6) $ symmetries are broken to flavor $ U(1) $
symmetries. Three such symmetries arise from the subgroup of the
observable $ E_{8} $ which is orthogonal to $ SO(10) $. Additional 
$ U(1) $ symmetries arise from the pairing of real fermions.  The final
observable gauge group depends on the number of such pairings.

%%%%%%%%%%%%%%%%%%%%%%%%%%%%%%%%%%%%%%%%%%%%%%%%%%%%%%%%%%%%%%%%%%%%%%%%%%
\subsection{$ \mathbb{Z}_{2} \times \mathbb{Z}_{2} $ correspondence}

The precise geometrical realization of the full $ N_{\textrm{gen}} = 3 $
models is not yet known.  However, the extended NAHE set
$ \{ \mathbf{1}, \mathbf{S}, \mathbf{b}_{1}, \mathbf{b}_{2},
     \mathbf{b}_{3}, \boldsymbol{\xi}_{1} \} $,
or equivalently
$ \{ \mathbf{1}, \mathbf{S}, \boldsymbol{\xi}_{1}, \boldsymbol{\xi}_{2},
\mathbf{b}_{1}, \mathbf{b}_{2} \} $, has been shown to yield the same data
as the $ \mathbb{Z}_{2} \times \mathbb{Z}_{2} $ orbifold of a
toroidal Narain model \cite{Nar:New} with nontrivial background fields
\cite{NarSarWit:A_No}. This Narain model has
$ \mathcal{N} = 4 $ spacetime supersymmetry and either
\begin{equation}
SO(12) \times E_{8} \times E_{8}
\end{equation}
or
\begin{equation}
SO(12) \times SO(16) \times SO(16)
\end{equation}
gauge group, depending on the
choice of sign for the GSO projection coefficient
$ C (^{ \boldsymbol{\xi_{1}} }_{ \boldsymbol{\xi_{2}} } ) $.
Let the former and latter Narain models be denoted by
$ \mathbf{N}_{+} $ and $ \mathbf{N}_{-} $, respectively.
The corresponding $ \mathbb{Z}_{2} \times \mathbb{Z}_{2} $ orbifolds are
\begin{equation}
\mathbf{Z}_{+}
  \equiv \frac{ \mathbf{N}_{+} }
              { \mathbb{Z}_{2} \times \mathbb{Z}_{2} },
\quad
\mathbf{Z}_{-}
  \equiv \frac{ \mathbf{N}_{-} }
              { \mathbb{Z}_{2} \times \mathbb{Z}_{2} }.
\end{equation}
In the free-fermionic formulation, the
$ \mathbf{N}_{+} $ and $ \mathbf{N}_{-} $  
data is produced by the set
$\{ \mathbf{1}, \mathbf{S}, \boldsymbol{\xi}_{1}, \boldsymbol{\xi}_{2} \}$
with appropriate choices for the sign of
$ C (^{ \boldsymbol{\xi_{1}} }_{ \boldsymbol{\xi_{2}} } ) $.
Adding the basis vectors $ \mathbf{b}_{1} $ and $ \mathbf{b}_{2} $
corresponds to $ \mathbb{Z}_{2} \times \mathbb{Z}_{2} $ modding.  These
vectors reduce the spacetime supersymmetry to $ \mathcal{N} = 1 $, break
\begin{equation}
SO(12) \rightarrow SO(4)^{3}
\end{equation}
and either
\begin{equation}
E_{8} \times E_{8} \rightarrow E_{6} \times U(1)^{2} \times E_{8}
\end{equation}
or
\begin{equation}
SO(16) \times SO(16) \rightarrow SO(10) \times U(1)^{3} \times SO(16).
\end{equation}
The three sectors
\begin{equation*}
\mathbf{b}_{1} \oplus ( \mathbf{b}_{1} + \boldsymbol{\xi}_{1} ), \quad
\mathbf{b}_{2} \oplus ( \mathbf{b}_{2} + \boldsymbol{\xi}_{2} ), \quad
\mathbf{b}_{3} \oplus ( \mathbf{b}_{3} + \boldsymbol{\xi}_{3} )
\end{equation*}
correspond to the three twisted sectors of the
$ \mathbb{Z}_{2} \times \mathbb{Z}_{2} $ orbifold, each sector producing
eight generations in either the
$ \mathbf{27} $ representation of $ E_{6} $ or
$ \mathbf{16} $ representation of $ SO(10) $.
The
\begin{equation*}
\textrm{Neveu-Schwarz} \, \oplus \, \boldsymbol{\xi}_{1}
\end{equation*} 
sector corresponds to
the untwisted sector, producing an additional three
$ \mathbf{27} $ and $ \mathbf{ \overline{27} } $ or
$ \mathbf{16} $ and $ \mathbf{ \overline{16} } $
repesentations of $ E_{6} $ or $ SO(10) $, respectively.

As the NAHE set is common to all realistic free-fermionic
models, $ \mathbf{Z}_{+} $ and $ \mathbf{Z}_{-} $ are at their core.  
To make a connection with the Calabi-Yau 3-folds $ \mathbf{Z} $ and
$ \mathbf{X} $, we construct the 
$ \mathbb{Z}_{2} \times \mathbb{Z}_{2} $ orbifold at a generic point in
the Narain moduli space.  Start with the 10-dimensional space compactified
on the torus $ \mathbf{T}^{2}_{1} \times \mathbf{T}^{2}_{2} \times
\mathbf{T}^{2}_{3} $
parameterized by three complex cooordinates
$ z_{1} $, $ z_{2} $, and $ z_{3} $, with the identification
\begin{equation}
z_{i} = z_{i} + 1, \quad
z_{i} = z_{i} + \tau_{i} \quad \quad
(i = 1,2,3)
\end{equation}
where $ \tau_{i} $ is the complex parameter of the torus
$ \mathbf{T}^{2}_{i} $.  Under a  $ \mathbb{Z}_{2} $ twist
$ z_{i} \rightarrow - z_{i} $, the torus $ \mathbf{T}^{2}_{i} $
has four fixed points at
\begin{equation}
z_{i} = \{ 0, 1/2, \tau_{i}/2, (1 + \tau_{i})/2 \}.
\end{equation}
With the two $ \mathbb{Z}_{2} $ twists
\begin{align}
\alpha: \ & ( z_{1},z_{2},z_{3} ) \rightarrow ( -z_{1}, - z_{2},  z_{3} )
\\
\beta: \ & ( z_{1},z_{2},z_{3} ) \rightarrow (  z_{1}, - z_{2}, - z_{3} )
\end{align}
there are three twisted sectors, $ \alpha $, $ \beta $, and
$ \alpha \beta  = \alpha \cdot \beta $, each with 16 fixed points
and producing 16 generations in the $ \mathbf{27} $ representation of
$ E_{6} $, for a total of 48.  The untwisted sector
produces an additional three 
$ \mathbf{27} $ and $ \mathbf{\overline{27}} $
representations of $ E_{6} $. Denote this orbifold by $ \mathbf{X}_{+} $.
It corresponds to an elliptically fibered Calabi-Yau 3-fold 
$ \mathbf{X} $ with $ (h^{(1,1)},h^{(2,1)}) = (51,3) $ \cite{Voi-Bor}.
Now, add the shift
\begin{equation}
\gamma : ( z_{1},z_{2},z_{3} ) \rightarrow
         \left(
           z_{1} + \frac{1}{2} , z_{2} + \frac{1}{2}, z_{3} + \frac{1}{2}
         \right).
\end{equation}
The product of the $ \gamma $ shift with any of the three twisted sectors
$ \alpha $, $ \beta $, and $ \alpha \beta $ does not produce any
additional fixed points.  Therefore, $ \gamma $ acts freely.  Under the
action of $ \gamma $, the fixed points from the twisted sectors are
identified in pairs.  Hence, each twisted sector now has 8 fixed points
and produces 8 generations in the $ \mathbf{27} $ representation of
$ E_{6} $, for a total of 24. Consequently, this orbifold has
$ ( h^{(1,1)}, h^{(2,1)} ) = (27,3) $, reproducing
the data of $ \mathbf{Z}_{+} $.  Since $ \gamma $ acts freely, the    
associated Calabi-Yau 3-fold has $ \pi_{1} = \mathbb{Z}_{2} $
\cite{BerEllFarNanQiu:Tow}. Thus, $ \mathbf{Z}_{+} $ and $ \mathbf{X}_{+}
$ correspond to the Calabi-Yau 3-folds $ \mathbf{Z} $ and $ \mathbf{X} $,
respectively.

%%%%%%%%%%%%%%%%%%%%%%%%%%%%%%%%%%%%%%%%%%%%%%%%%%%%%%%%%%%%%%%%%%%%%%%%%%
\subsection{Nonperturbative top quark Yukawa coupling}

The Yukawa coupling of the top quark is obtained at the cubic level of the
superpotential and is a coupling between states from the
twisted-twisted-untwisted sectors.  For example, in the Standard-like
models \cite{(SLM)}, the relevant coupling is
$ t^{c}_{1} Q_{1} \overline{h}_{1} $, where $ t^{c}_{1} $ and $ Q_{1} $
are respectively the quark SU(2) singlet and doublet from the sector
$ \mathbf{b}_{1} $, and $ \overline{h}_{1} $ is the untwisted Higgs.
One can calculate this coupling in the full $ N_{\textrm{gen}} = 3 $
model, or at the level of either the
$ \mathbf{Z}_{+} $ or $ \mathbf{Z}_{-} $ orbifolds as a
$ \mathbf{27}^{3} $ $ E_{6} $
or
$ \mathbf{16} \cdot \mathbf{16} \cdot \mathbf{10} $  $ SO(10) $ coupling,
respectively. The nonperturbative top quark Yukawa coupling at the grand  
unification scale is computed, at least in principle, as an integral over
$ \mathbf{Z} $ \cite{LukOvrWal:Fiv}, with $ \mathbf{Z} $ taken
to be the Calabi-Yau 3-fold associated with $ \mathbf{Z}_{+} $ or
$ \mathbf{Z}_{-} $.

%%%%%%%%%%%%%%%%%%%%%%%%%%%%%%%%%%%%%%%%%%%%%%%%%%%%%%%%%%%%%%%%%%%%%%%%%%
%%%%%%%%%%%%%%%%%%%%%%%%%%%%%%%%%%%%%%%%%%%%%%%%%%%%%%%%%%%%%%%%%%%%%%%%%%
\section{\label{SumRul}Summary of rules}

This section summarizes rules for constructing heterotic M-theory vacua
with $ H = E_{6} $, $ SO(10) $, and $ SU(5) $ grand unification groups and
arbitrary $ N_{\textrm{gen}} $.  For more details, refer to
\cite{FarGar:Yuk} or \cite{thesis}.  These rules are an
adaptation of those presented in \cite{DonOvrPanWal}.
Compactification to four dimensions
with $ \mathcal{N} = 1 $ supersymmetry is achieved on a torus-fibered
Calabi-Yau 3-fold $ \mathbf{Z} = \mathbf{X} / \tau_{\mathbf{X}} $
with first homotopy group $ \pi_{1}(\mathbf{Z}) = \mathbb{Z}_{2} $.
The $ H = E_{6} $, $ SO(10) $, and $ SU(5) $ vacua correspond
\cite{Don-Uhl-Yau} to semistable holomorphic vector bundles 
$ \mathbf{V}^{ \textrm{\tiny{(1)}} }_{\mathbf{Z}} $ over $ \mathbf{Z} $
having structure group
$ G_{\mathbb{C}} = SU(n)_{\mathbb{C}} $ with $ n = 3 $, 4, and 5,
respectively.  The vector bundles 
$ \mathbf{V}^{ \textrm{\tiny{(1)}} }_{\mathbf{Z}} $ and
$ \mathbf{V}^{ \textrm{\tiny{(2)}} }_{\mathbf{Z}} $      
are located on the observable and hidden orbifold fixed planes,
$ \mathbf{M}^{10}_{ \textrm{\tiny{(1)}} } $ and 
$ \mathbf{M}^{10}_{ \textrm{\tiny{(2)}} } $,
respectively.  $ \mathbf{V}^{ \textrm{\tiny{(2)}} }_{\mathbf{Z}} $
is taken to be a trivial bundle so that $ E_{8} $ remains unbroken in the
hidden sector.  The vacua with nonstandard embeddings may contain 
$ N $ M5-branes in the bulk space which wrap holomorphic curves 
$ W^{(\ell)}_{\mathbf{Z}} $ $ (\ell = 1,\ldots,N) $ in $ \mathbf{Z} $.
The class of the total M5-brane curve is denoted by $ [W_{\mathbf{Z}}] $.
$ \beta^{(0)}_{i} $ and $ \beta^{(N+1)}_{i} $ 
$ ( i=1,\ldots,h^{(1,1)}(\mathbf{Z}) ) $ are the instanton charges on the
observable and hidden orbifold fixed planes, respectively, and 
$ \beta^{(\ell)}_{i} $ are the magnetic charges of the M5-branes. 
Requiring vanishing instanton charges in the observable sector yields 
potentially viable matter Yukawa couplings \cite{ArnDut:Yuk}.  In a nutshell, 
the procedure is

\begin{enumerate}
\item Construct $ \mathbf{Z} = \mathbf{X} / \tau_{\mathbf{X}} $.

\item Construct $ \mathbf{V}^{\textrm{\tiny{(1)}}}_{\mathbf{Z}} $ over $
\mathbf{Z} $
with structure group $ G_{\mathbb{C}} = SU(n)_{\mathbb{C}} $.

\item $ N_{\textrm{gen}} 
  = \frac{1}{2} \int_{\mathbf{Z}} c_{3}
(V^{ \textrm{\tiny{(1)}} }_{\mathbf{Z}}) $

\item Require $ [W_{\mathbf{Z}}]
  =   c_{2}(\mathbf{TZ})
    - c_{2} \! \left( \mathbf{V}^{ \textrm{\tiny{(1)}} }_{\mathbf{Z}}
\right)
    - c_{2} \! \left( \mathbf{V}^{ \textrm{\tiny{(2)}} }_{\mathbf{Z}}
\right) $ for anomaly cancellation.  

\item Require $ \beta^{(0)}_{i} = 0 $.

\end{enumerate}

The elliptically fibered Calabi-Yau 3-fold 
$ \mathbf{X} \stackrel{\pi}{\rightarrow} \mathbf{B} $ admits global
sections $ \sigma $ and $ \xi $ which satisfy
$ \xi + \xi = \sigma $.
This condition facilitates the construction of the freely acting
involution $ \tau_{\mathbf{X}} $ as the composition
$ \tau_{\mathbf{X}} = \alpha \circ t_{\xi} $.
Here, $ \alpha $ is the lift to $ \mathbf{X} $ of an involution 
$ \tau_{\mathbf{B}} $ on the base $ \mathbf{B} $, and
$ t_{\xi}(x) = x + \xi(b) \ (x \in \pi^{-1}(b), \ b \in \mathbf{B}) $
is a translation on the fiber $ \pi^{-1}(b) $. The Calabi-Yau condition
$ c_{1}(\mathbf{TX}) = 0 $ restricts the base \cite{restrictB} to
be a del Pezzo $ (\mathbf{dP}_{r}, r = 0,\dots,8) $, rational elliptic
$ (\mathbf{dP}_{9}) $, Hirzebruch $ (\mathbb{F}_{r}, r \geq 0) $, blown-up
Hirzebruch, or an Enriques surface.

A semistable holomorphic vector bundle 
$ \mathbf{V}^{ \textrm{\tiny{(1)}} }_{\mathbf{X}} $ over
$ \mathbf{X} $ with structure group 
$ G_{\mathbb{C}} = SU(n)_{\mathbb{C}} $
can be constructed by the
\emph{spectral cover method} \cite{SpeCovMet}. 
This method requires the specification of a divisor $ \mathbf{C} $ of 
$ \mathbf{X} $ (known as the \emph{spectral cover}) and a line bundle 
$ \boldsymbol{\mathcal{N}} $ over $ \mathbf{C} $.  The condition
$ c_{1}(\mathbf{V}^{ \textrm{\tiny{(1)}} }_{\mathbf{X}}) = 0 $
implies that the spectral data
$ (\mathbf{C},\boldsymbol{\mathcal{N}}) $
can be written in terms of an effective divisor class $ \eta $ in the base
$ \mathbf{B} $ and coefficients $ \lambda $ and $ \kappa_{i} $
$ (i = 1, \ldots, 4 \eta \cdot c_{1}(\mathbf{B})) $.  Constraints are
placed on $ \eta $ , $ \lambda $ , and the $ \kappa_{i} $ by the condition
that
\begin{equation}
c_{1}(\boldsymbol{\mathcal{N}}) = n \left( \frac{1}{2} + \lambda \right)
\sigma
   + \left( \frac{1}{2} - \lambda \right) \pi^{*}_{\mathbf{C}} \eta
   + \left( \frac{1}{2} + n \lambda \right) \pi^{*}_{\mathbf{C}}
c_{1}(\mathbf{B})
   + \textstyle{\sum_{i}} \kappa_{i} N_{i}
\end{equation}
be an integer class.  To ensure that $ c_{1}(\boldsymbol{\mathcal{N}}) $
is an integer class, one can impose the sufficient (but not necessary)
Class A or Class B constraints discussed in Section \ref{Vacua}. More
generally, $ c_{1}(\boldsymbol{\mathcal{N}}) $ will be an integer class if
the constraints
\begin{gather}
\label{BCsigma}
q \equiv n \left( \frac{1}{2} + \lambda \right) \in \mathbb{Z}
\\
\label{BCetaChern}
\left(\frac{1}{2} - \lambda \right) \pi^{*}_{\mathbf{C}} \eta
   + \left( \frac{1}{2} + n \lambda \right) \pi^{*}_{\mathbf{C}}
                                            c_{1}(\mathbf{B})
\quad \textrm{is an integer class}
\\
\label{BCkappa}
\kappa_{i} - \frac{1}{2} m \in \mathbb{Z}, \quad m \in \mathbb{Z}
\end{gather}
are simultaneously satisfied.  To ensure that $ H $ is the largest
subgroup of $ E_{8} $ preserved by 
$ \mathbf{V}^{ \textrm{\tiny{(1)}} }_{\mathbf{X}} $, impose the stability
constraint \cite{BerMay:Sta} $ \eta \geq n c_{1}(\mathbf{B}) $ 
$ (n \geq 2) $.  Requiring 
$ \tau^{*}_{\mathbf{X}}(\mathbf{V}^{ \textrm{\tiny{(1)}} }_{\mathbf{X}}) 
= \mathbf{V}^{ \textrm{\tiny{(1)}} }_{\mathbf{X}} $  ensures that 
$ \mathbf{V}^{ \textrm{\tiny{(1)}} }_{\mathbf{X}} $ descends to 
a bundle $ \mathbf{V}^{ \textrm{\tiny{(1)}} }_{\mathbf{Z}} $ over 
$ \mathbf{Z} $.

%%%%%%%%%%%%%%%%%%%%%%%%%%%%%%%%%%%%%%%%%%%%%%%%%%%%%%%%%%%%%%%%%%%%%%%%%%
%%%%%%%%%%%%%%%%%%%%%%%%%%%%%%%%%%%%%%%%%%%%%%%%%%%%%%%%%%%%%%%%%%%%%%%%%%
\section{\label{Vacua}$ N_{\textrm{gen}} = 3 $, $ H=SO(10) $ vacua}

$ N_{\textrm{gen}} = 3 $, $ H=SO(10) $ heterotic M-theory vacua
with nonvanishing and vanishing instanton charges in the observable sector
are searched for in \cite{FarGarIsi:Non} and \cite{FarGar:Yuk},
respectively.  In the former work, the search is restricted to vector
bundles satisfying the Class A or Class B constraints:
\begin{align}
\label{ClassA}
\textrm{Class A:}&  \quad \lambda \in \mathbb{Z}, \quad
\eta = c_{1}(\mathbf{B}) \ \textrm{mod} \ 2, \quad
\kappa_{i} - \frac{1}{2} m \in \mathbb{Z}
\\
\label{ClassB}
\textrm{Class B:}&  \quad \lambda - \frac{1}{2} \in \mathbb{Z}, \quad
c_{1}(\mathbf{B}) \ \textrm{is an even class}, \quad
\kappa_{i} - \frac{1}{2} m \in \mathbb{Z}
\end{align}
where $ m \in \mathbb{Z} $.  In the latter work, the more general
constraints (\ref{BCsigma}), (\ref{BCetaChern}), and 
(\ref{BCkappa}) with $ n = 4 $ are imposed.  Table \ref{vacuasum}
summarizes the results.
\begin{table}[t]

$$ \begin{array}{ll}
\beta^{(0)}_{i} \neq 0 & \exists \ \mathbf{B} = \mathbb{F}_{2} \
\textrm{Class B vacua}
\\
                       & \exists \ \mathbf{B} = \mathbb{F}_{r} \
(r \ \textrm{even} \geq 4) \ \textrm{Class B vacua*}
\\
                       & \nexists \ \mathbf{B} = \mathbb{F}_{0} \
\textrm{Class A or Class B vacua}
\\
                       & \nexists \ \mathbf{B} = \mathbb{F}_{r} \
(r \ \textrm{odd} \geq 1) \ \textrm{Class A or Class B vacua}
\\
                       & \nexists \ \mathbf{B} = \mathbf{dP}_{3} \
\textrm{Class A or Class B vacua}
\\
\\
\beta^{(0)}_{i} = 0   & \exists \ \mathbf{B} = \mathbf{dP}_{7}
\ \textrm{vacua*}
\\
                      & \nexists \ \mathbf{B} = \mathbf{dP}_{r} \
(r = 0,\ldots,6,8) \ \textrm{vacua}
  
\\
                      & \nexists \ \mathbf{B} = \mathbb{F}_{r} \
(r \geq 0) \ \textrm{vacua}
\end{array} $$

\caption{$ H = SO(10) $ heterotic M-theory vacua with
$ N_{\textrm{gen}} = 3 $.  The `$*$' indicates that these vacua have been
demonstrated to exist when certain constraints on $ \tau_{\mathbf{B}} $
are satisfied.}
\label{vacuasum}
\end{table}

%%%%%%%%%%%%%%%%%%%%%%%%%%%%%%%%%%%%%%%%%%%%%%%%%%%%%%%%%%%%%%%%%%%%%%%%%%
%%%%%%%%%%%%%%%%%%%%%%%%%%%%%%%%%%%%%%%%%%%%%%%%%%%%%%%%%%%%%%%%%%%%%%%%%%
\section*{Acknowledgements}

I would like to thank Alon E. Faraggi and Jose M. Isidro.

%%%%%%%%%%%%%%%%%%%%%%%%%%%%%%%%%%%%%%%%%%%%%%%%%%%%%%%%%%%%%%%%%%%%%%%%%%
%%%%%%%%%%%%%%%%%%%%%%%%%%%%%%%%%%%%%%%%%%%%%%%%%%%%%%%%%%%%%%%%%%%%%%%%%%
%%%%%%%%%%%%%%%%%%%%%%%%%%%%%%%%%%%%%%%%%%%%%%%%%%%%%%%%%%%%%%%%%%%%%%%%%%   

\end{document}